\documentclass[cameraready]{Interspeech}
\usepackage{cite}

\title{Phonetic Error Analysis of Raw Waveform Acoustic Models}

\author[affiliation={1,2}]{Erfan}{Loweimi}
\author[affiliation={3}]{Zhengjun}{Yue}
\author[affiliation={1}]{Andrea}{Carmantini}
\author[affiliation={4}]{Zoran}{Cvetkovic}
\author[affiliation={1}]{Steve}{Renals}
\author[affiliation={1}]{Peter}{Bell}

\address{
  $^1$Centre for Speech Technology Research (CSTR), University of Edinburgh, UK\\
  $^2$Cisco, UK;
  $^3$SLAI \& CUHK-SZ, China;
  $^4$King's College London, UK
  }
\email{}

\newcommand{\specialcell}[2][c]{%
\begin{tabular}[#1]{@{}c@{}}#2\end{tabular}}

\begin{document}

\maketitle

\begin{abstract}
We analyse error patterns of raw waveform acoustic models on TIMIT phone recognition beyond the overall phone error rate (PER). PER is decomposed across three broad phonetic class (BPC) categorisations, and confusion matrices are constructed from substitution errors.
Our models combine parametric (SincNet, Sinc2Net) or non-parametric CNNs with Bidirectional LSTMs, achieving 13.9\%/15.3\% PER on Dev/Test, the best reported results for raw waveform models on TIMIT. Transfer learning from WSJ reduces PER to 11.3\%/12.3\%, surpassing the Filterbank baseline.
Per-BPC analysis reveals that BLSTM layers benefit transition-dependent classes most, while WSJ transfer learning improves consonants roughly three times more than vowels. Confusion patterns are consistent across raw waveform and Filterbank systems, indicating that the dominant confusions reflect inherent phonetic similarities.
\end{abstract}

\section{Introduction}
\label{sec:intro}

The conventional metric for evaluating phone recognition systems is phone error rate (PER), measuring the Levenshtein distance \cite{levenshtein1966binary} involving substitution, deletion, and insertion errors. While PER is a useful aggregate metric, it provides no insight into which broad phonetic classes (BPCs) contribute most to the errors or how different classes are confused with one another.

In~\cite{lea2023}, a detailed phonetic error analysis was conducted for Filterbank-based, GMM-HMM and DNN-HMM systems. Phones were grouped into three categorisations: an 8-class set \{affricate, diphthong, fricative, nasal, plosive, semi-vowel, vowel, silence\}, a 3-class consonant/vowel$^+$/silence set, and a 3-class voiced/unvoiced/silence set. Substitution, deletion, insertion, and overall PER were computed for each BPC, with confusion matrices constructed from substitution errors to reveal systematic confusion patterns.

This paper extends the phonetic error analysis of Filterbank-based systems in~\cite{lea2023} to raw waveform acoustic models \cite{sainath2015learning}. Unlike Filterbank-based systems, raw waveform models jointly learn speech parameterisation with the acoustic model, avoiding potentially lossy feature engineering while retaining access to Fourier phase spectrum information, which has proven useful across various tasks \cite{INTERSPEECH2011,choi2019phase,ICASSP2017,Loweimi2020RawSignMagnitude,yue23_interspeech,INTERSPEECH2017_ph,Myphdthesis,ICASSP2021}. Whether these models exhibit similar per-BPC error distributions and confusion patterns to conventional systems, or whether their learnable front-end yields qualitatively different behaviour, remains unexplored.

The main contributions of this work are as follows:
\begin{itemize}
    \item We present raw waveform acoustic models combining parametric (SincNet~\cite{sincnet1}, Sinc2Net~\cite{INTERSPEECH2019}) or non-parametric CNNs with BLSTMs~\cite{Graves2005}, achieving the best reported PERs for raw waveform models trained from scratch on TIMIT~\cite{TIMIT}.

    \item We provide a per-BPC breakdown of PER, substitution, deletion and insertion errors, along with confusion matrices for three phonetic categorisations.

    \item We analyse the effect of BLSTM layers on each BPC, showing that classes with strong temporal dynamics (Diphthongs, Fricatives, Semi-vowels) benefit most.

    \item We examine the per-BPC impact of WSJ~\cite{WSJ} transfer learning, revealing a consistent consonant--vowel gain asymmetry and comparing with the Filterbank-based baseline.

    \item We compare per-BPC error distributions and confusion patterns between raw waveform and Filterbank systems.
\end{itemize}

Having reviewed the related work in Section~2, we define the phonetic categorisations in Section~3. Section~4 describes the acoustic model architecture, and Section~5 presents the results along with discussion. Section~6 concludes the paper.

\section{Related Work}
\label{sec:review}
Raw waveform models with conventional (non-parametric) CNNs have been widely applied for phone recognition~\cite{PALAZ201915,Yousafzai11a,Ager2011}, large-vocabulary speech recognition~\cite{Tuske2018}, multi-channel beamforming~\cite{sainath2016ICASSP}, speaker adaptation~\cite{Ghahremani2016}, and multi-scale representations~\cite{ZhuIS2016,Platen2019}.
A complementary line of research replaces non-parametric convolution layers with parameterised ones defined by a small number of learnable parameters. SincNet~\cite{sincnet1} employs sinc-function kernels corresponding to rectangular filters in the frequency domain, where each filter is specified solely by its centre frequency and bandwidth. It has been applied to phone recognition~\cite{sincnet1,e2e-sincnet2020}, hybrid~\cite{INTERSPEECH2020-wave} and end-to-end~\cite{e2e-sincnet2020} speech recognition, and speaker recognition~\cite{sincnet1}.
In~\cite{INTERSPEECH2019}, this approach was generalised to modulated kernel-based CNNs, specifically Sinc2Net, GammaNet, and GaussNet, which use triangular, Gammatone, and Gaussian frequency responses, respectively. Other parametric architectures include ParzNet~\cite{Oglic2021} and Complex Gabor CNNs~\cite{cgcnn2020}. Parametric CNNs have also been applied to dysarthric speech recognition~\cite{INTERSPEECH2022} and speaker adaptation via filter retraining~\cite{FeinburgASRU2019}.
Beyond phonetic error analysis, BPCs have been applied in speaker verification~\cite{Eatock1994,Margit2006}, speech enhancement~\cite{Lu2020}, language identification~\cite{Kempton2008}, speech coding~\cite{Zhang1997}, emotion recognition~\cite{Yuan2021TheRO}, and ASR tasks such as state clustering~\cite{Young2006} and multilingual recognition~\cite{ZGANK2005379}.

\section{Phonetic Categorisations}
\label{sec:bpc}

We use three phonetic categorisations following~\cite{lea2023}, specified in Table~\ref{tab:bpc}. Silence encompasses non-speech segments, epenthetic silence~\cite{TIMIT}, short pauses, and closures before plosives. Vowel$^+$ denotes the union of vowels and diphthongs, which are grouped due to their acoustic similarity~\cite{lea2023}.

The overall phone error rate (PER) is decomposed across broad phonetic classes (BPCs) as
\begin{equation}
\text{PER} = \sum_{c \in C} \text{PER}_c,
\end{equation}

\noindent where $C$ denotes the set of all BPCs in a given categorisation and $c \in C$ represents an individual class. For example, in the voiced/unvoiced/silence categorisation, $C = \{\text{Voiced}, \text{Unvoiced}, \text{Silence}\}$, and $\text{PER}_c$ corresponds to the phone error rate associated with each of these classes.

\section{Raw Waveform Acoustic Model}
\label{sec:arch}
\begin{figure}[t]
  \centering
  \includegraphics[width=\linewidth]{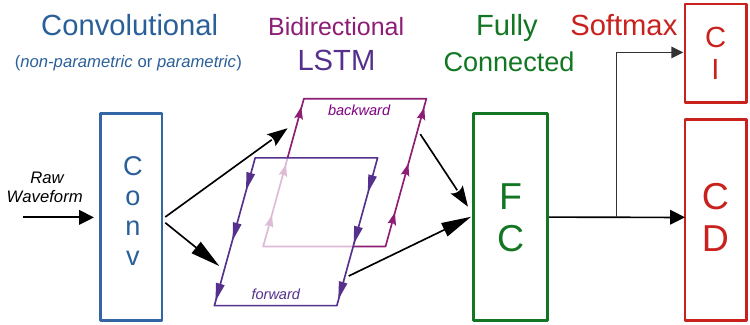}
  \caption{Proposed raw waveform acoustic model architecture. The convolutional layer (parametric or non-parametric) is followed by BLSTM and fully-connected (FC) layers, with context-dependent (CD) and context-independent (CI) output heads.}
  \label{fig:arch}
\end{figure}

Fig.~\ref{fig:arch} depicts the architecture of our acoustic model: a cascade of (parametric or non-parametric) convolutional, Bidirectional LSTM (BLSTM)~\cite{Graves2005}, and fully-connected (FC) layers. The CNN extracts spectral features, the BLSTM captures temporal context, and the FC layer improves class separability before the softmax classifier. The output layer has two heads: context-dependent (CD) state-clustered triphones (primary) and context-independent (CI) monophones (regularisation).

We experimented with both parametric and non-parametric convolutional layers. For the parametric variant, we use SincNet~\cite{sincnet1}, whose filters are rectangular in the frequency domain, and Sinc2Net~\cite{INTERSPEECH2019}, whose Sinc-squared kernel yields triangular filters closely comparable with the Mel Filterbank (FBank).

Both CNN and FC sub-networks contain one layer. The FC layer has 1024 nodes; the convolutional layer has 128 kernels of length 129 with max pooling of size~4. Dropout~\cite{Srivastava2014} and ReLU are used in both. The BLSTM layers contain 550 nodes per direction with dropout and batch normalisation~\cite{batchnorm2015}.

\section{Experimental Results and Discussion}
\label{sec:exp}

Models were trained using PyTorch-Kaldi~\cite{pytorch-kaldi,Kaldi2011,pytorch2017} with cross-entropy loss and a batch size of 8. The CD and CI output heads consist of 1936 and 48 nodes, respectively. Reference FBank features are 83-D (80 filters plus three pitch-related features). For transfer learning, systems were pre-trained on WSJ; subsequently, only the weights between the penultimate and output layers were re-initialised and trained from scratch on TIMIT.

\begin{table}[t!]
\centering
\caption{Phone-to-BPC mapping for the three categorisations.}
\vspace{-2mm}
\begin{tabular}{l|c}
\hline
         Classes & Phones  \\
\hline
\hline
\noalign{\vskip 0mm}
Affricates (Aff)  & ch jh  \\
Diphthongs (Dip)  & aw ay ey ow oy \\
Fricatives (Fri)  & dh f s sh th v z \\
Nasals (Nas)      & m n ng \\
Plosives (Plo)    & b d dx g k p t \\
Semi-vowels (Sem) & hh l r w y \\
Silence (Sil)     & sil \\
Vowels (Vow)      & aa ae ah eh er ih iy uh uw \\
\hline
Consonants (Con)    & \specialcell{b ch d dh dx f g hh jh k l m n ng p r s sh\\t th v w y z}  \\
Silence (Sil)       & sil \\
Vowel$^+$ (Vow$^+$) & aw ay ey ow oy aa ae ah eh er ih iy uh uw \\
\hline
Voiced (Voi)    & \specialcell{aa ae ah aw ay b d dh dx eh er ey g hh\\ih iy jh l m n ng ow oy r uh uw v w y z} \\
Silence (Sil)   & sil \\
Unvoiced (Unv)  & ch f k p s sh t th \\
\hline
\hline
\end{tabular}
\label{tab:bpc}
\end{table}

\begin{table}[t!]
\centering
\caption{PERs (\%) of raw waveform systems on TIMIT.}
\vspace{-2mm}
\begin{tabular}{l|c|cc}
\hline
 Feature  & Architecture & Dev & Test  \\
\hline
\hline
FBank-83~\cite{lea2023} & Best System in~\cite{lea2023} & 12.8 & 14.1  \\
\hline
Raw-Wav~\cite{PALAZ201915} & CNN & - & 21.9 \\
Raw-Wav (E2E)~\cite{e2e-sincnet2020}  & CNN & 18.9 & 21.1  \\
Raw-Wav (E2E)~\cite{e2e-sincnet2020}  & SincNet & 17.3 & 19.3  \\
Raw-Wav~\cite{sincnet1}  & CNN & - & 18.1 \\
Raw-Wav~\cite{sincnet1}  & SincNet & - & 17.2  \\
Raw-Wav~\cite{INTERSPEECH2019} & GammaNet & - & 17.2 \\
Raw-Wav~\cite{cgcnn2020}  & CGCNN & 15.2 & 17.1  \\
Raw-Wav~\cite{INTERSPEECH2019} & GaussNet & - & 17.0 \\
Raw-Wav~\cite{INTERSPEECH2019} & Sinc2Net & - & 16.9 \\
Raw-Wav~\cite{Oglic2021} & ParzNet & 15.0 & 16.5 \\
Raw-Wav~\cite{Loweimi2023-real} & CNN & 14.9 & 16.5 \\
\hline
Raw-Wav--Proposed & CNN+BLSTM & 13.9 & 15.8  \\
Raw-Wav--Proposed & SincNet+BLSTM & 14.2 & 15.6 \\
Raw-Wav--Proposed & Sinc2Net+BLSTM & 13.9 & 15.3 \\
\hline
\hline
\end{tabular}
\label{tab:per}
\end{table}

\begin{table}[t!]
  \centering
  \caption{PERs (\%) on TIMIT with WSJ transfer learning.}
  \vspace{-1mm}
  \begin{tabular}{l|c|cc}
  \hline
   Feature  & Architecture & Dev & Test  \\
  \hline
  \hline
  FBank-83-WSJ~\cite{lea2023} & Best System in~\cite{lea2023} & 11.5 & 13.1 \\
  \hline
  Raw-Wav--Proposed & CNN+BLSTM & 11.3 & 12.3 \\
  Raw-Wav--Proposed & SincNet+BLSTM & 11.3 & 12.5 \\
  Raw-Wav--Proposed & Sinc2Net+BLSTM & 11.5 & 12.6 \\
  \hline
  \hline
  \end{tabular}
  \label{tab:per-wsj}
  \end{table}

Table~\ref{tab:per} compares the PER of various raw waveform systems on TIMIT. Our proposed models, combining parametric (SincNet, Sinc2Net) and non-parametric (CNN) first layers with BLSTMs, outperform all prior raw waveform systems. Sinc2Net+BLSTM achieves the lowest Test PER of 15.3\% among models trained solely on TIMIT. With WSJ transfer learning (Table~\ref{tab:per-wsj}), all three proposed models surpass the FBank-WSJ system from~\cite{lea2023}, with CNN+BLSTM reaching 11.3\%/12.3\% on Dev/Test.

\subsection{Effect of Sequential Modelling}
\label{ssec:lstm}

To isolate the contribution of BLSTM layers, we compare CNN-only models (three convolutional layers, no recurrence) with the full CNN+BLSTM architecture for all three front-ends. Table~\ref{tab:lstm} reports the relative PER reduction per BPC.

\begin{table}[t!]
\centering
\caption{Relative PER reduction (\%) per BPC after adding BLSTM layers. D: TIMIT Dev, T: TIMIT Test. Higher values indicate greater benefit from sequential modelling.}
\vspace{-1mm}
{\small
\begin{tabular}{l|cc|cc|cc}
\hline
& \multicolumn{2}{c|}{CNN} & \multicolumn{2}{c|}{SincNet} & \multicolumn{2}{c}{Sinc2Net} \\
BPC & D & T & D & T & D & T \\
\hline
\hline
\noalign{\vskip 0mm}
Aff & $-$11.1 & $-$22.2 & $-$7.1 & 0.0 & 0.0 & $-$66.7 \\
Dip & 33.6 & 28.0 & 28.9 & 25.3 & 20.5 & 27.8 \\
Fri & 24.0 & 22.6 & 16.8 & 20.3 & 17.1 & 11.0 \\
Nas & 11.9 & 10.7 &  3.9 &  6.6 & 11.8 & 16.8 \\
Plo & 16.1 & 14.9 &  5.0 & $-$4.9 &  8.7 & 14.0 \\
Sem & 19.3 &  6.1 & 14.9 & 15.2 & 19.0 &  9.0 \\
Sil & 13.3 &  2.1 & 18.6 &  6.4 &  1.6 & $-$8.0 \\
Vow & 14.2 &  7.3 &  5.9 &  5.1 &  9.7 &  6.2 \\
\hline
Con & 18.2 & 13.5 & 10.8 & 9.4 & 14.3 & 11.1 \\
Vow$^+$ & 16.4 &  9.9 &  8.6 & 7.7 & 10.9 &  8.9 \\
\hline
Voi & 17.4 & 11.3 &  9.8 & 9.8 & 12.8 & 10.3 \\
Unv & 16.9 & 14.9 &  9.1 & 0.0 & 11.0 &  7.8 \\
\hline
\hline
\end{tabular}}
\label{tab:lstm}
\end{table}

Three BPCs stand out with consistently large gains across all three convolutional front-ends: Diphthongs (28\% average), Fricatives (19\%), and Semi-vowels (18\%). This consistency across both parametric and non-parametric convolutions suggests a general property of these phonetic classes rather than an artefact of a specific front-end. The shared characteristic of these three classes is their reliance on temporal dynamics: Diphthongs are defined by formant trajectories gliding between two vowel targets; Semi-vowels (glides and liquids) are similarly characterised by rapid spectral transitions; Fricatives require sustained spectral context to distinguish acoustically similar sounds (e.g., /s/ vs /z/). Convolutional layers, operating on fixed-length local windows, capture spectral snapshots but cannot model these multi-frame dependencies. The BLSTM layers, by encoding the full temporal context bidirectionally, resolve the ambiguity inherent in such transition-dependent classes.

Conversely, Vowels gain only $\sim$10\% on average, consistent with their relatively stationary spectral structure: vowel identity is largely determined by the formant pattern within a single analysis frame. Affricates show no reliable gain, likely due to the small sample size (only two phones: /ch/ and /jh/).

The 3-class categorisations corroborate these findings: Consonants and Vowel$^+$ both improve on Dev and Test, as do the Voiced and Unvoiced classes. The Test set exhibits the same trends with higher variance, as expected from the smaller evaluation set. These results confirm that adding sequential modelling to raw waveform systems provides the greatest benefit for BPCs whose discrimination depends on temporal context.

\subsection{Effect of Transfer Learning}
\label{ssec:tl}

Having established the role of sequential modelling, we now examine the effect of additional training data via WSJ transfer learning. Table~\ref{tab:tl} reports the relative PER reduction per BPC.

\begin{table}[t!]
\centering
\caption{Relative PER reduction (\%) per BPC after WSJ transfer learning. D: TIMIT Dev, T: TIMIT Test. Higher values indicate greater benefit from additional training data.}
\vspace{-1mm}
{\small
\begin{tabular}{l|cc|cc|cc}
\hline
& \multicolumn{2}{c|}{CNN} & \multicolumn{2}{c|}{SincNet} & \multicolumn{2}{c}{Sinc2Net} \\
BPC & D & T & D & T & D & T \\
\hline
\hline
\noalign{\vskip 0mm}
Aff & 56.7 & 18.2 & 56.7 & $-$27.3 & 56.0 & 46.7 \\
Dip & 20.2 &  7.4 & 24.0 & 16.1 & 32.7 &  7.7 \\
Fri & 30.1 & 31.4 & 31.2 & 23.8 & 40.6 & 31.2 \\
Nas & 46.6 & 49.0 & 50.3 & 31.3 & 29.6 & 13.1 \\
Plo & 15.3 & 30.0 & 24.6 & 33.9 & 14.1 & 22.7 \\
Sem & 33.3 & 36.8 & 30.8 & 31.0 & 26.8 & 33.1 \\
Sil &  0.0 &  0.0 & $-$19.3 & $-$27.3 &  7.2 &  5.6 \\
Vow &  7.9 &  9.1 & 11.0 & 10.6 &  8.9 & 10.4 \\
\hline
Con & 30.5 & 35.3 & 32.8 & 29.2 & 28.3 & 27.0 \\
Vow$^+$ &  9.0 &  8.9 & 12.2 & 11.1 & 11.3 & 10.1 \\
\hline
Voi & 19.4 & 21.8 & 21.1 & 19.8 & 18.8 & 17.2 \\
Unv & 22.0 & 25.5 & 30.9 & 23.1 & 25.1 & 27.7 \\
\hline
\hline
\end{tabular}}
\label{tab:tl}
\end{table}

The dominant trend is a consonant--vowel asymmetry: Consonants improve by $\sim$30\% on average while Vowel$^+$ gains only $\sim$10\%, and this 3:1 ratio is remarkably stable across both evaluation sets and all three front-ends. Within the 8-class breakdown, Nasals, Fricatives and Semi-vowels consistently show large gains. These classes share a key property: their acoustic realisation is strongly shaped by phonetic context, whether through coarticulation (Nasals), context-dependent spectral variation (Fricatives) or rapid formant transitions (Semi-vowels). A larger training corpus provides more diverse phonetic environments, directly benefiting recognition of such context-dependent sounds.

Vowels gain only $\sim$10\% on both sets. Vowel identity is primarily governed by formant frequencies, which reflect vocal tract geometry and are heavily speaker-dependent~\cite{Eatock1994,Margit2006}. Since WSJ SI-284 provides more data but from fewer speakers (284 vs TIMIT's 630), it does not proportionally expand the speaker diversity needed for vowel discrimination. Silence shows near-zero or negative gain, as its acoustic realisation is largely shaped by recording conditions, limiting the utility of cross-corpus transfer. Affricates vary widely between sets due to the very small class size. While individual 8-class BPCs show some variance between Dev and Test (e.g., Diphthongs and Plosives), the 3-class categorisations are highly stable, confirming that the consonant--vowel asymmetry is a stable finding.

\subsection{Confusion Patterns}
\label{ssec:confusion}

\begin{figure}[t!]
  \centering
  \includegraphics[width=\linewidth,height=116mm]{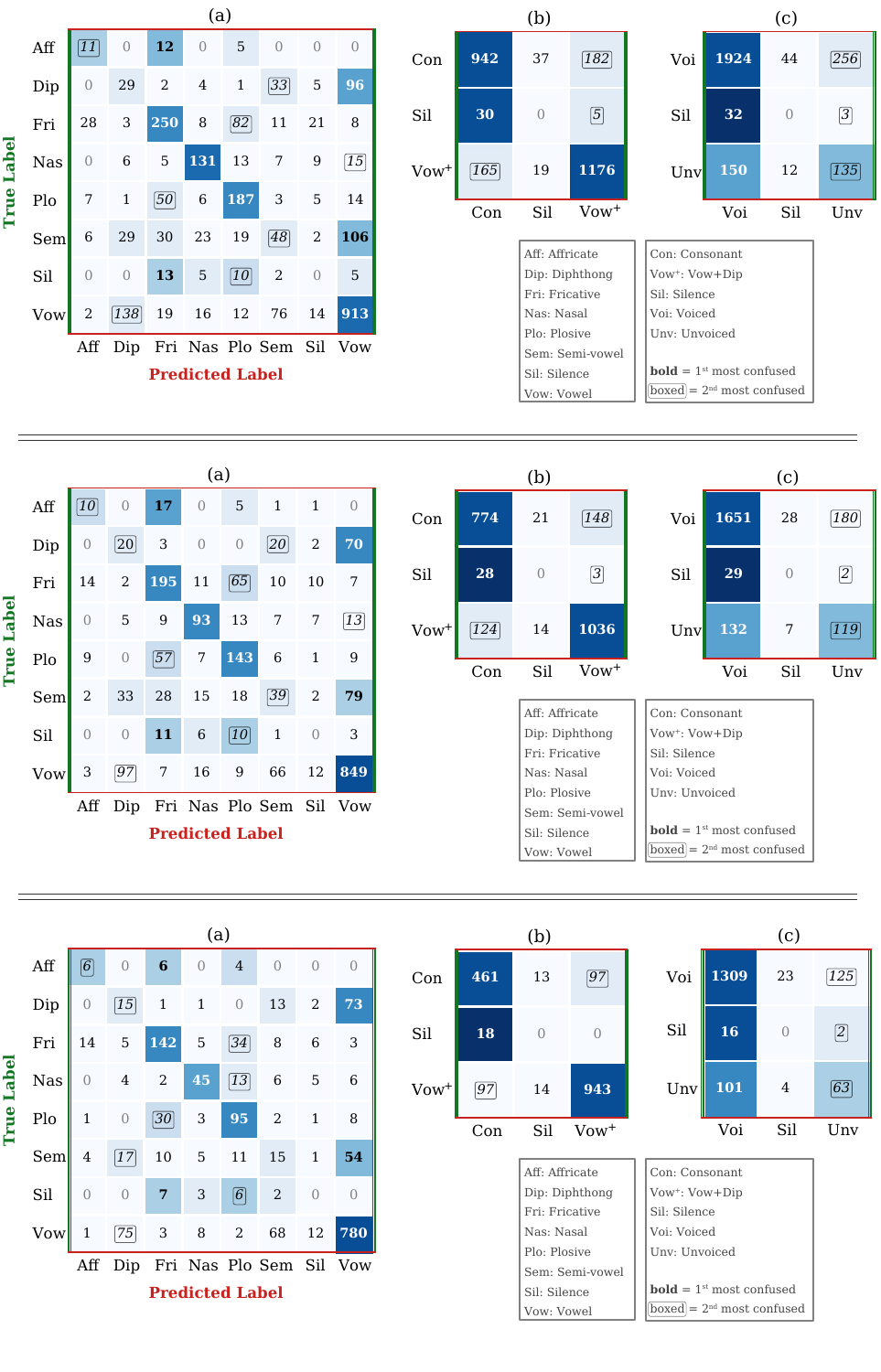}
  \vspace{-7mm}
  \caption{BPC confusion matrices aggregated over Dev and Test sets. Top: CNN (no BLSTM); middle: CNN+BLSTM; bottom: CNN+BLSTM with WSJ transfer learning. Columns: (a)~8-class, (b)~Consonant/Vowel$^+$/Silence, (c)~Voiced/Unvoiced/Silence. Darker shading indicates higher row-normalised substitution percentage. \textbf{Bold}: most confused class per row; \textit{boxed}: second most confused.}
  \label{fig:conf}
  \vspace{-1mm}
\end{figure}

Fig.~\ref{fig:conf} presents confusion matrices computed from substitution errors aggregated over TIMIT’s Dev and Test sets for three models: the CNN without BLSTM layers (top row), the full CNN+BLSTM (middle), and CNN+BLSTM with WSJ transfer learning (bottom). Entry $[i,j]$ counts how often phones in BPC~$i$ are substituted by phones in BPC~$j$. The Silence diagonal is zero because Silence contains a single phone (Table~\ref{tab:bpc}).

Two clusters of confusable classes persist across all three models: Plosives~$\leftrightarrow$~Fricatives and Vowels~$\leftrightarrow$~Diphthongs~$\leftrightarrow$~Semi-vowels. These clusters reflect phonetic proximity: Plosives and Fricatives share place-of-articulation cues and both involve noise-like spectral energy, while Diphthongs, Semi-vowels and Vowels all involve a relatively open vocal tract and are distinguished primarily by formant trajectories. These patterns match the FBank confusion matrices reported in~\cite{lea2023}, confirming that the dominant confusions stem from inherent phonetic similarities rather than system-specific artefacts.

\textbf{Effect of sequential modelling on confusion.}
Comparing the top and middle rows of Fig.~\ref{fig:conf}, adding BLSTM layers reduces the total substitution count but preserves the confusion structure entirely: the 1st and 2nd most confused classes remain identical for every BPC in all three categorisations. In the Consonant/Vowel$^+$/Silence view, within-class confusion rates rise from 81\% to 82\% for Consonants and from 87\% to 88\% for Vowel$^+$, showing that BLSTM resolves proportionally more cross-class confusions. Among the 8-class BPCs, Vowels show the clearest within-class improvement (77\%~$\to$~80\%), consistent with BLSTM's ability to model formant trajectories that distinguish vowels from acoustically overlapping Diphthongs and Semi-vowels. The Voiced within-class rate rises from 87\% to 89\%, while the Unvoiced rate remains at~46\%.

\textbf{Effect of transfer learning on confusion.}
Comparing the middle and bottom rows of Fig.~\ref{fig:conf}, WSJ transfer learning further reduces substitution errors while preserving the 1st most confused class for all BPCs. Three 8-class BPCs show a shift in the 2nd most confused class. Nasals shift from Vowels to Plosives (16\% of Nasal substitutions become Plosives, up from 9\%), reflecting the known perceptual similarity between nasal and plosive closures at shared places of articulation. Diphthongs and Semi-vowels both shift their 2nd confusion target to Diphthongs, as transfer learning concentrates their errors more narrowly within the vowel-like cluster. The Vowel$^+$ within-class rate rises from 88\% to 90\%, while the Consonant rate stays at~81\%, confirming that transfer learning refines the confusion distribution without introducing new confusion patterns.

\subsection{Filterbank vs Raw Waveform}
\label{ssec:fbank-raw}

\begin{figure}[t!]
  \centering
  \includegraphics[width=0.95\linewidth,height=56.5mm]{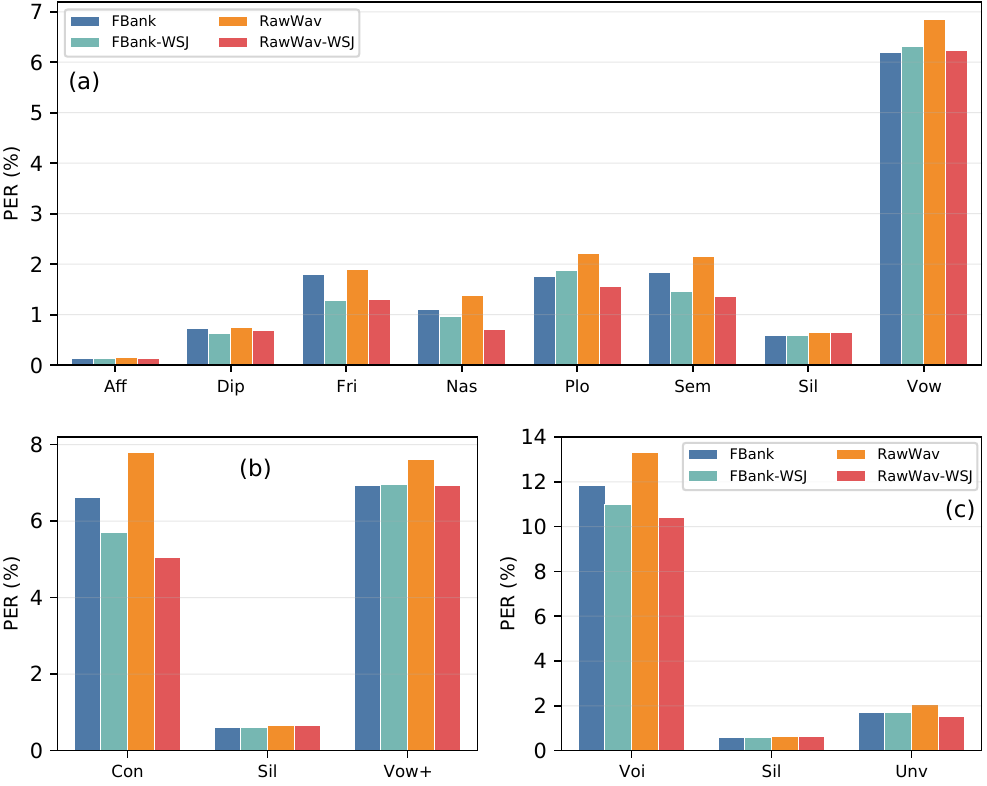}
  % \vspace{-1mm}
  \caption{Per-BPC PER (\%) comparison between FBank and raw waveform CNN models on TIMIT Test, with and without WSJ transfer learning. (a)~8-class BPCs, (b)~Consonant/Vowel$^+$/Silence, (c)~Voiced/Unvoiced/Silence.}
  \label{fig:fbank-raw}
\end{figure}

Finally, we compare the raw waveform CNN with the FBank baseline. Both systems share the same BLSTM back-end, so any performance difference is attributable to the front-end: fixed magnitude-based features vs.\ a learnable convolutional layer operating on raw waveforms. Fig.~\ref{fig:fbank-raw} shows the per-BPC PER for both models, with and without WSJ transfer learning.

Without transfer learning, FBank achieves lower PER than the raw waveform CNN on nearly every BPC, as the fixed pipeline provides a stronger inductive bias when data is limited. With WSJ transfer learning, the raw waveform CNN surpasses FBank on most classes. The largest improvements occur in Nasals, Fricatives, and Semi-vowels. This reversal confirms that raw waveform models have higher data requirements but, once met, the jointly learned representation outperforms the fixed magnitude-based pipeline.

Comparing the raw waveform CNN confusion matrices (Fig.~\ref{fig:conf}) with the FBank system from~\cite{lea2023}, the overall confusion structure is similar: both exhibit the Plosive$\leftrightarrow$Fricative and Vowel$\leftrightarrow$Diphthong$\leftrightarrow$Semi-vowel clusters. Silence is confused predominantly with consonantal classes in both systems, as plosive closures and fricative noise share the low-energy, aperiodic characteristics of silence segments. This overall similarity indicates that the dominant confusions are governed by inherent acoustic-phonetic similarities between classes rather than the choice of feature representation.
%Within this shared structure, differences do emerge: the raw waveform CNN achieves a higher within-class confusion rate for Nasals than FBank, suggesting the learnable front-end captures nasal resonances more effectively, and the primary confusion target for Silence differs (Fricatives for the raw waveform CNN vs Plosives for FBank).
After WSJ transfer learning, a larger proportion of Silence substitutions fall into consonantal classes in both systems, as the additional data improves discrimination among speech sounds but does not resolve the Silence$\leftrightarrow$Plosive and Silence$\leftrightarrow$Fricative confusions.%, where plosive closures and weak fricative segments are acoustically close to silence. 
~For Vowel$^+$, both systems show minimal gain from transfer learning, consistent with the speaker-dependent nature of vowel identity~\cite{Eatock1994,Margit2006}, as WSJ SI-284 introduces more training data but fewer speakers (284) than TIMIT (630).
%In the voiced/unvoiced/silence grouping, silence is overwhelmingly confused with voiced rather than unvoiced classes despite both lacking fundamental frequency. This is largely an artefact of the BPC mapping: voiced plosive closures (bcl, dcl, gcl), which carry voice bar energy, are grouped under Silence and when substituted naturally become their corresponding voiced plosives (b, d, g), inflating the Silence$\to$Voiced count. Class priors (voiced phones constitute the majority of frames) further amplify this bias.

\section{Conclusion}
We analysed per-BPC error distributions and confusion patterns of raw waveform acoustic models on TIMIT, achieving the best reported raw waveform PERs (11.3\%/12.3\% Dev/Test with WSJ transfer learning), outperforming the Filterbank baseline. Confusion patterns are consistent across raw waveform and Filterbank systems, reflecting inherent phonetic similarities. BLSTM layers benefit transition-dependent classes most, while WSJ transfer learning improves consonants roughly three times more than vowels across all front-ends. These per-BPC insights can guide targeted improvements such as class-specific data augmentation or loss weighting; extending this analysis to end-to-end and self-supervised models is a natural next step.

\clearpage
\bibliographystyle{IEEEtran}
{\small \bibliography{ref}}

@INPROCEEDINGS{Ager2011,
author={Ager, M. and Cvetković, Z. and Sollich, P.},
booktitle={{ISIT}}, 
title={Combined waveform-cepstral representation for robust speech recognition}, 
year={2011},
volume={},
number={},
pages={864-868},
doi={10.1109/ISIT.2011.6034260}}

@article{Yousafzai11a,
  title={Combined Features and Kernel Design for Noise Robust Phoneme Classification Using Support Vector Machines},
  author={J. Yousafzai and P. Sollich and Z. Cvetkovic and B. Yu},
  journal = {IEEE/ACM Trans. Audio, Speech and Lang. Proc.},
  volume = {19},
  pages = {1396--1407},
  year={2011}
 }

@article{Oglic2021,
author    = {Dino Oglic and
           Zoran Cvetkovic and
           Peter Sollich},
title     = {Learning Waveform-Based Acoustic Models Using Deep Variational Convolutional
           Neural Networks},
journal = {IEEE/ACM Trans. Audio, Speech and Lang. Proc.},
volume    = {29},
pages     = {2850--2863},
year      = {2021},
doi       = {10.1109/TASLP.2021.3104193},
}

@inproceedings{INTERSPEECH2011,
  author = {Loweimi, E. and Ahadi, S.M. and Sheikhzadeh, H.},
  booktitle = {INTERSPEECH},
  title = {Phase-Only Speech Reconstruction Using Very Short Frames.},
  year = {2011}
}

@INPROCEEDINGS{ICASSP2021,
author={Loweimi, E. and Cvetkovic, Z. and Bell, P. and Renals, S.},
booktitle={ICASSP},
title={Speech Acoustic Modelling from Raw Phase Spectrum},
year={2021}
}

@article{Srivastava2014,
  author  = {N. Srivastava and others},
  title   = {Dropout: A Simple Way to Prevent Neural Networks from Overfitting},
  journal = {Journal of Machine Learning Research},
  year    = {2014},
  volume  = {15},
  pages   = {1929-1958},
}

@book{Young2006,
  author = {Young, S.J. and others},
  publisher = {Cambridge University Press},
  title = {{The HTK Book Version 3.4}},
  year = 2006
}

@INPROCEEDINGS{Kaldi2011,
    author={Povey, D. and others},
    title = {The {Kaldi} Speech Recognition Toolkit},
    booktitle = {ASRU},
    year = {2011},
}

@inproceedings{INTERSPEECH2022,
  author={Zhengjun Yue and Erfan Loweimi and Heidi Christensen and Jon Barker and Zoran Cvetkovic},
  title={{Dysarthric Speech Recognition From Raw Waveform with Parametric CNNs}},
  year=2022,
  booktitle={{INTERSPEECH}},
  pages={31--35},
  doi={10.21437/Interspeech.2022-163}
}

@inproceedings{INTERSPEECH2017_ph,
author               = {Loweimi, E. and  Barker, J. and Saz Torralba, O. and Hain, T. },
booktitle            = {INTERSPEECH},
title                = {Robust Source-filter Separation of Speech Signal in the Phase Domain},
year                 = {2017},
pages                = {414-418},
}

@INPROCEEDINGS{ICASSP2017,
author={Loweimi, E. and Barker, J. and Hain, T.},
booktitle={ICASSP},
title={Statistical normalisation of phase-based feature representation for robust speech recognition},
year={2017},
pages={5310-5314},
doi={10.1109/ICASSP.2017.7953170},
}

@inproceedings{Graves2005,
  title={Bidirectional LSTM Networks for Improved Phoneme Classification and Recognition},
  author={Alex Graves and Santiago Fern{\'a}ndez and J{\"u}rgen Schmidhuber},
  booktitle={ICANN},
  year={2005}
}

@phdthesis{Myphdthesis,
 author = {Loweimi, Erfan},
 title = {Robust Phase-based Speech Signal Processing; From Source-Filter Separation to Model-Based Robust ASR},
 year = {2018},
 school = {University of Sheffield},
 url = {http://etheses.whiterose.ac.uk/19409/},
}

@inproceedings{pytorch2017,
  title={Automatic differentiation in {PyTorch}},
  author={Paszke, A. and Gross, S. and Chintala, S. and Chanan, G. and Yang, E.d and DeVito, Z. and Lin, Z. and Desmaison, A. and Antiga, L. and Lerer, A.},
  booktitle={{NIPS} Workshop on Autodiff},
  year={2017}
}

@inproceedings{pytorch-kaldi,
    author = {Ravanelli, M. and Parcollet, T. and Bengio, Y.},
    title = {The {PyTorch-Kaldi} Speech Recognition Toolkit},
    year = {2019},
    booktitle = {ICASSP}
}

@inproceedings{sincnet1,
    author = {Ravanelli, M. and Bengio, Y.},
    title  = {Speaker and Speech Recognition from Raw Waveform with {SincNet}}, 
    booktitle  = {{ICASSP}},
      year = {2019}
}

@INPROCEEDINGS{cgcnn2020,
  author={P. {Noé} and T. {Parcollet} and M. {Morchid}},
  booktitle={ICASSP},
  title={{CGCNN}: Complex Gabor Convolutional Neural Network on Raw Speech}, 
  year={2020},
  volume={},
  number={},
  pages={7724-7728},}

@INPROCEEDINGS{e2e-sincnet2020,
  author={T. {Parcollet} and M. {Morchid} and G. {Linarès}},
  booktitle={ICASSP}, 
  title={{E2E-SINCNET}: Toward Fully End-To-End Speech Recognition}, 
  year={2020},
  volume={},
  number={},
  pages={7714-7718},}

@misc{TIMIT,
    author = {Garofolo, J. S. and Lamel, L. F. and Fisher, W. M. and Fiscus, J. G. and Pallett, D. S. and Dahlgren, N. L.},
    publisher = {NIST},
    title = {{DARPA} {TIMIT} Acoustic Phonetic Continuous Speech Corpus},
    year = {1993}
}

@inproceedings{WSJ,
  author = {Paul, D. B. and Baker, J. M.},
  booktitle = {ICASSP},
  pages = {899--902},
  title = {The Design for the {Wall Street Journal-based CSR} Corpus},
  year = 1992
}

@article{PALAZ201915,
title = {End-to-end acoustic modeling using convolutional neural networks for HMM-based automatic speech recognition},
journal = {Speech Communication},
volume = {108},
pages = {15-32},
year = {2019},
issn = {0167-6393},
doi = {https://doi.org/10.1016/j.specom.2019.01.004},
author = {Dimitri Palaz and Mathew Magimai-Doss and Ronan Collobert},
}

@INPROCEEDINGS{sainath2016ICASSP,
  author={Sainath, Tara N. and Weiss, Ron J. and Wilson, Kevin W. and Narayanan, Arun and Bacchiani, Michiel},
  booktitle={{ICASSP}}, 
  title={Factored spatial and spectral multichannel raw waveform {CLDNN}s}, 
  year={2016},
  volume={},
  number={},
  pages={5075-5079},
  doi={10.1109/ICASSP.2016.7472644}}

@InProceedings {Tuske2018,
author= {T\"uske, Z. and Schl\"uter, R. and Ney, H.},
title= {Acoustic modeling of Speech Waveform based on Multi-Resolution, Neural Network Signal Processing},
booktitle= {{ICASSP}},
year= {2018},
}

@InProceedings {ZhuIS2016,
author= {Zhu, Z. and Engel, J. H. and Hannun, A.},
title= {Learning Multiscale Features Directly From Waveforms},
booktitle= {{INTERSPEECH}},
year= 2016,
}

@inproceedings{Ghahremani2016,
  title={Acoustic Modelling from the Signal Domain Using {CNN}s},
  author={Ghahremani, P. and Manohar, V. and Povey, D. and Khudanpur, S.},
  booktitle={{INTERSPEECH}},
  year={2016}
}

@InProceedings {Platen2019,
author= {von Platen, P. and Zhang, C. and Woodland, P. C.},
title= {Multi-Span Acoustic Modelling using Raw Waveform Signals},
booktitle= {{INTERSPEECH}},
year= 2019,
}

@inproceedings{INTERSPEECH2019,
  title     = "On Learning Interpretable {CNN}s with Parametric Modulated Kernel-based Filters",
  author    = "Loweimi, E. and Bell, P. and Renals, S.",
  year      = "2019",
booktitle= {{INTERSPEECH}},
}

@inproceedings{INTERSPEECH2020-wave,
  author={E. Loweimi and P. Bell and S. Renals},
  title={{On the Robustness and Training Dynamics of Raw Waveform Models}},
  year=2020,
  booktitle={{INTERSPEECH}},
  pages={1001--1005},
  doi={10.21437/Interspeech.2020-0017},
}

@inproceedings{FeinburgASRU2019,
  title     = {Acoustic model adaptation from raw waveforms with {SincNet}},
  author    = {Fainberg, J. and Klejch, O. and Loweimi, E. and Bell, P. and Renals, S.},
  year      = {2019},
booktitle= {{ASRU}},
}

@inproceedings{batchnorm2015,
 author = {Ioffe, S. and Szegedy, C.},
 title = {Batch Normalization: Accelerating Deep Network Training by Reducing Internal Covariate Shift},
 booktitle = {{ICML}},
 year = {2015},
}

@INPROCEEDINGS{Eatock1994,
  author={Eatock, J.P. and Mason, J.S.},
  booktitle={{ICASSP}}, 
  title={A quantitative assessment of the relative speaker discriminating properties of phonemes}, 
  year={1994},
  pages={133--136},
  doi={10.1109/ICASSP.1994.389337}
  }

@INPROCEEDINGS{Zhang1997,
  author={Lei Zhang and Tian Wang and Cuperman, V.},
  booktitle={{ICASSP}}, 
  title={A {CELP} variable rate speech codec with low average rate}, 
  year={1997},
  volume={2},
  number={},
  pages={735-738},
  doi={10.1109/ICASSP.1997.596022}}

@inproceedings{Margit2006,
author = {Antal, Margit and Toderean, Gavril},
title = {Speaker Recognition and Broad Phonetic Groups},
booktitle = {{SPPRA}},
year = {2006},
isbn = {0889865809},
publisher = {ACTA Press},
pages = {155–159},
series = {SPPRA'06}
}

@article{Yuan2021TheRO,
  title={The Role of Phonetic Units in Speech Emotion Recognition},
  author={Jiahong Yuan and Xingyu Cai and Renjie Zheng and Liang Huang and Kenneth Church},
  journal={ArXiv},
  year={2021},
  volume={abs/2108.01132}
}

@article{ZGANK2005379,
title = {Data-driven generation of phonetic broad classes, based on phoneme confusion matrix similarity},
journal = {Speech Communication},
volume = {47},
number = {3},
pages = {379-393},
year = {2005},
issn = {0167-6393},
doi = {https://doi.org/10.1016/j.specom.2005.03.011},
author = {Andrej Žgank and Bogomir Horvat and Zdravko Kačič},
}

@inproceedings{Lu2020,
  author={Yen-Ju Lu and Chien-Feng Liao and Xugang Lu and Jeih-weih Hung and Yu Tsao},
  title={{Incorporating Broad Phonetic Information for Speech Enhancement}},
  year=2020,
  booktitle={{INTERSPEECH}},
  pages={2417--2421},
}

@inproceedings{Kempton2008,
  author    = {Timothy Kempton and
               Roger K. Moore},
  title     = {Language identification: insights from the classification of hand annotated phone transcripts},
  booktitle = {Odyssey},
  publisher = {{ISCA}},
  year      = {2008},
}

@ARTICLE{Loweimi2023-real,
  author={Loweimi, Erfan and Yue, Zhengjun and Bell, Peter and Renals, Steve and Cvetkovic, Zoran},
  journal={IEEE/ACM Transactions on Audio, Speech, and Language Processing}, 
  title={Multi-Stream Acoustic Modelling Using Raw Real and Imaginary Parts of the Fourier Transform}, 
  year={2023},
  volume={31},
  number={},
  pages={876-890},
  doi={10.1109/TASLP.2023.3237167}}

@ARTICLE{lea2023,
  author={Loweimi, Erfan and Carmantini, Andrea and Bell, Peter and Renals, Steve and Cvetkovic, Zoran},
  journal={IEEE/ACM Transactions on Audio, Speech, and Language Processing}, 
  title={Phonetic Error Analysis Beyond Phone Error Rate}, 
  year={2023},
  volume={31},
  number={},
  pages={3346-3361},
  doi={10.1109/TASLP.2023.3313417}}

@inproceedings{sainath2015learning,
  title={Learning the Speech Front-End With Raw Waveform {CLDNNs}},
  author={Sainath, Tara N. and others},
  booktitle            = {{INTERSPEECH}},
  year={2015},
  doi={10.21437/Interspeech.2015-1}
}

@inproceedings{Loweimi2020RawSignMagnitude,
  author    = {Erfan Loweimi and Peter Bell and Steve Renals},
  title     = {Raw Sign and Magnitude Spectra for Multi-Head Acoustic Modelling},
  booktitle = {{INTERSPEECH}},
  pages     = {1644--1648},
  year      = {2020},
  doi       = {10.21437/Interspeech.2020-18}
}

@inproceedings{choi2019phase,
  title={Phase-Aware Speech Enhancement with Deep Complex U-Net},
  author={Choi, Hyeong-Seok and Kim, Jang-Hyun and Huh, Jaesung and Kim, Adrian and Ha, Jung-Woo and Lee, Kyogu},
  booktitle={{ICLR}},
  year={2019}
}

@article{levenshtein1966binary,
  title={Binary codes capable of correcting deletions, insertions and reversals},
  author={Levenshtein, Vladimir I.},
  journal={Soviet Physics Doklady},
  volume={10},
  number={8},
  pages={707--710},
  year={1966}
}

@inproceedings{yue23_interspeech,
  title     = {Dysarthric Speech Recognition, Detection and Classification using Raw Phase and Magnitude Spectra},
  author    = {Zhengjun Yue and Erfan Loweimi and Zoran Cvetkovic},
  year      = {2023},
  booktitle={{INTERSPEECH}},
  pages     = {1533--1537},
  doi       = {10.21437/Interspeech.2023-222},
  issn      = {2958-1796},
}

\end{document}